# Review on Microscopic Pedestrian Simulation Model

Kardi Teknomo[1], Yasushi Takeyama[2], Hajime Inamura[3]

Microscopic Pedestrian Simulation Model is computer simulation model of pedestrian movement where every pedestrian in the model is treated as individual. Most of pedestrian researches have been done on macroscopic level [for best classical examples: Fruin (1971), HCM (1985)], which does not consider the interaction between pedestrians and does not well suited for prediction of pedestrian flow performance in pedestrian areas or building with some objects that reduce the effective width of it. In the other hand, microscopic level has more general usage and considers detail of the design. Tough the analytical model for microscopic pedestrian model is existed exist [Henderson (1974), Helbing (1992)], the numerical solution of the model is very difficult and simulation is favorable. The model has practical application of Evacuation from building, Design of pedestrian area, and Experimental & Optimization Design Tool. There are several microscopic pedestrian simulation models:

**a. Benefit Cost Cellular Model**
Gipps and Marksjo (1985) propose this model. It simulates the pedestrian as particle in a cell. Each cell can be occupied by at most one pedestrian and a score assigned to each cell on the basis of proximity to pedestrians. The score represent the gain made by the pedestrian when moving toward his destination. The repulsive effect of the nearby pedestrians, and formulated as:

$$\text{Score} = \frac{K \cdot (S_i - X_i) \cdot (D_i - X_i) \cdot |(S_i - X_i) \cdot (D_i - X_i)|}{|S_i - X_i|^2 \cdot |D_i - X_i|^2} - \frac{1}{(\Delta - \alpha)^2 + \beta} \quad (1)$$

Where the field of two pedestrian overlap, the score in each cell is the sum of the score generated by pedestrian individually. The score is calculated in the nine-cell neighbor of the pedestrian (including the location of the pedestrian). Pedestrian will move to the next cell that has maximum net benefit.

**b. Magnetic Force Model**
Prof. Okazaki (1979-93) developed this model with Matsushita. The application of magnetic models and equation of motion in the magnetic field cause pedestrian movement. Each pedestrian and obstacle have positive pole. Negative pole is assumed to be located at the goal of pedestrians. Pedestrian moves to their goals and avoids collisions. Two forces are work on each pedestrian. First, magnetic force as formulated by Coulomb's law, which is depend on the intensity of magnetic load of a pedestrian and distance between pedestrians. Another force acts on a pedestrian to avoid the collision with another pedestrian or obstacle exerts acceleration **a** is calculated as:

$\mathbf{a} = \mathbf{V} \cdot \cos(\alpha) \cdot \tan(\beta)$ (2)

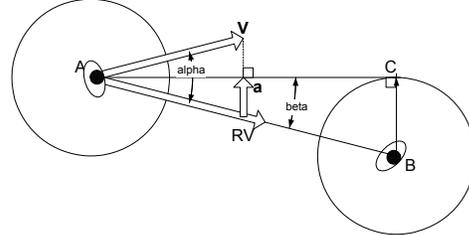

Figure 1. Additional Repulsive Force on Magnetic Force Model

Total of forces from goals, walls and other pedestrians act on each pedestrian, and its decides the velocity of each pedestrian each time.

**c. Social Force Model**
Helbing (1991-99) was developed Social Force Model with Molnar, and Vicsek which has similar principles of both Benefit Cost cellular Model and Magnetic Force Model. A pedestrian is subjected to social forces that motivate the pedestrian. The summation of these forces act upon a pedestrian create acceleration dv/dt as:

$$m\frac{d\vec{v}_i(t)}{dt} = m\frac{v_o\vec{e}_i - \vec{v}_i(t) + \vec{\xi}_i(t)}{\tau} + \sum_{j(\neq i)} \vec{f}_{ij}(\vec{x}_i(t), \vec{x}_j(t)) + \vec{f}_b(\vec{x}_i(t)) \quad (3)$$

The first term in the right hand of eq.(3) represent the motivation to reach the goal. The model based on assumption that every pedestrian has intention to reach certain destination at a certain target time. The direction is a unit vector from a particular location to the destination point.

---

[1] Doctoral Student, Graduate School of Information Science, Tohoku University Japan
[2] Associate Professor, Graduate School of Information Science, Tohoku University Japan
[3] Professor, Graduate School of Information Science, Tohoku University Japan

**Table 1. Comparison Microscopic Pedestrian Simulation Models**

|  | **Benefit Cost Cellular** | **Magnetic Force** | **Social Force** |
|---|---|---|---|
| **Movement to goal** | Gain Score | Positive and negative magnetic force | Intended velocity |
| **Repulsive Effect** | Cost Score | Positive and positive magnetic forces | Interaction forces |
| **Pedestrian movement** | discreet | continuous | continuous |
| **Value of variables** | arbitrary | physical meaning | physical meaning |
| **Phenomena explained** | queuing | queuing, way finding in maze, evacuation | queuing, self - organization, oscillatory change |
| **Higher programming orientation in** | cellular based | heuristic | mathematics |
| **Evacuation Application** | possible | possible | not possible |
| **Parameter Calibration** | by inspection | by inspection | by inspection |

The ideal speed is equal to remaining distance per remaining time. The remaining distance is the different between the destination point and the location at that time, while the remaining time is different between target time and the simulation time. Intended Velocity is the ideal speed times the unit vector of direction. Speed limitation (maximum and minimum) can be put to make the speed more realistic. The second and last term of the right hand side of eq (3) designate for interaction between pedestrians and pedestrian to obstacles (i.e. column) and interaction pedestrian with the boundaries

**Comparison of the Models**

In general, Microscopic Pedestrian Simulation Model consist of two terms, that make the pedestrian moving toward the destination and make repulsive effect toward other pedestrian or obstacles. Benefit Cost cellular also uses arbitrary scores while Magnetic and Social Force model has more variables with physical meaning. Social Force model is highly mathematics evolved to explain the behavior of pedestrian, while Magnetic Force Model is more developed in heuristic approach.

None of the microscopic pedestrian simulation models have calibrated and validated their constant parameters based on real pedestrian movement data. It has no statistical guarantee that the parameters will work for general cases or even for specific region. Therefore, in order to use the model practically, future research direction is proposed to be done to automate data collection of individual movement data, and to set up statistical procedures for calibration and calibration of the parameters.

**Notation**

$S_i$ = vector location of target cell
$X_i$ = vector location of the subject
$D_i$ = vector location of destination
$\Delta$ = Distance between cell and the pedestrian.
$\alpha, \beta, K, m$ = Constants
**V** = velocity of pedestrian A
**RV** = relative velocity
$\mathbf{x}_i(t)$ = Location of pedestrian i at time t.
$\mathbf{v}_i(t)$ = velocity of pedestrian i at time t = $d\mathbf{x}_i(t)/dt$
$v_o$ = intended velocity.
$\mathbf{e}_i$ = direction of pedestrian i $\in \{(1,0),(=1,0)\}$
$\xi_i$ = Fluctuation of individual velocities